**Gene drive dynamics in natural populations: The importance of density-dependence, space and sex**

Sumit Dhole[1,2], Alun L. Lloyd[3,4,5] and Fred Gould[1,4,6]

1. Department of Entomology and Plant Pathology, North Carolina State University, Raleigh, NC, USA

2. ssdhole@ncsu.edu

3. Biomathematics Graduate Program and Department of Mathematics, North Carolina State University, Raleigh, NC, USA

4. Genetic Engineering and Society Center, North Carolina State University, Raleigh, NC, USA

5. alun_lloyd@ncsu.edu

6. fgould@ncsu.edu

**ABSTRACT:**

The spread of synthetic gene drives is often discussed in the context of panmictic populations connected by gene flow and described with simple deterministic models. Under such assumptions, an entire species could be altered by releasing a single individual carrying an invasive gene drive, such as a standard homing drive. While this remains a theoretical possibility, gene drive spread in natural populations is more complex and merits a more realistic assessment.

The fate of any gene drive released in a population would be inextricably linked to the population's ecology. Given the uncertainty often involved in ecological assessment of natural populations, understanding the sensitivity of gene drive spread to important ecological factors is critical. Here we review how different forms of density-dependence, spatial heterogeneity and mating behaviors can impact the spread of self-sustaining gene drives. We highlight specific aspects of gene drive dynamics and the target populations that need further research.

**INTRODUCTION:**

The last decade has seen an explosion of research on synthetic "gene drives" – genetic constructs that can spread through populations even when they decrease the viability and/or reproductive ability of individuals that carry them, much like naturally occurring selfish genetic elements (Burt & Trivers 2006). Recent interest in developing and assessing synthetic gene drive constructs has been fueled by their tremendous theoretical potential as tools for decreasing harm caused by vectors of diseases and other pests (Esvelt et al. 2014; Godfray et al. 2017; Gould 2008). For example, if genes that interfere with pathogen vectoring ability of a mosquito could be linked to a gene drive, both the drive and antipathogen genes would spread and would result in entire populations of vectors unable to transmit the pathogen (Curtis 1968). Gene drives could also be designed to drastically reduce fecundity or viability of drive-homozygous pest individuals, or cause extremely biased sex-ratios in the population (Craig et al. 1960), leading to suppression of the pest population as the drive spreads.

The idea of using genetic methods for controlling populations of pests has been around for at least eight decades (Curtis 1968; Serebrovsky 1940, 1969; Vanderplank 1947), but until recently, promising gene drive systems had eluded researchers. For much of the last century, genetic pest management research remained focused on using chromosomal translocations and transposable elements (reviewed by Gould & Schliekelman 2004). During the first decade of this century, research shifted to selfish genetic elements like meiotic-drivers, Medea, naturally occurring homing endonucleases (pre-CRISPR) and engineered underdominance (Macias et al. 2017; Sinkins & Gould 2006). Recent advances in genetic engineering technology, especially the advent of the CRISPR-Cas system, opened up the possibility of creating a host of new, highly potent gene drive systems (reviewed in Champer et al. 2016). While a number of new gene drive systems based on CRISPR-Cas remain simply as proposed designs (e.g. Champer et al. 2019a, 2020a; Dhole et al. 2019; Min et al. 2017; Noble et al. 2019; Prowse et al. 2017, 2019; Sudweeks et al. 2019), some have been built in yeast, insects, and mice and shown capable of driving through lab populations (Champer et al. 2018, 2019c, 2020b; DiCarlo et al. 2015; Grunwald et al. 2019; Hammond et al. 2016, 2018; Kyrou et al. 2018; Oberhofer et al. 2019a; Pham et al. 2019). Inference about the potential performance of such gene drives in wild populations has so far relied on mathematical models, lab experiments and on empirical assessments of natural selfish genetic elements (e.g. Cash et al. 2019). The general agreement between empirical data and even very simple models (Akbari et al. 2014; Buchman et al. 2018a,b; Champer et al. 2019c, 2020b; Kyrou et al. 2018; Oberhofer et al. 2019a; Schmidt et al. 2017; Windbichler et al. 2011) highlights the potential utility of models for anticipating the dynamics of these genetic elements.

Self-sustaining, synthetic gene drives work through one of two general mechanisms. The first type of gene drives biases the inheritance of the element in viable gametes from heterozygous parents. For example, meiotic drivers can propagate by disabling sperm of heterozygous males that don't contain the driver (Lindholm et al. 2016). Homing endonuclease gene (HEG) drives achieve a similar biased inheritance through conversion of wild-type alleles of heterozygous individuals into drive alleles (Godfray et al. 2017). The second general type of gene drives reduces the fitness of the wild-type alleles, and thus gains an advantage by biasing selection instead of inheritance. For example, natural and synthetic Medea elements spread by causing the death of either gametes or zygotes that lack them (Akbari et al. 2014; Beeman et al. 1992; Chen et al. 2007). Another example is gene drives based on simple, one-locus underdominance, which can cause reduced viability in heterozygotes. If an underdominant synthetic allele is more common in a population than the wild-type allele (e.g. due to a large release of transgenics), wild-type alleles suffer higher relative fitness loss than the synthetic alleles as they are more likely to form unfit heterozygotes (Sinkins & Gould 2006).

Gene drives can further be categorized in terms of other important properties besides the mechanism used to gain an advantage . Detailed descriptions of the general properties of a wide variety of gene drive designs have been compiled by other researchers (Burt 2014; Champer et al. 2016). Two fundamental criteria are 1) the intended effect of the gene drive on the population and 2) the frequency-dependent dynamics of different drives. The intended goal for "replacement drives" (also known as "modification drives") is to bring about a change in the genetic makeup of a natural population without significantly reducing the population size, as with a drive that spreads a pathogen-blocking gene in mosquitoes (e.g. Gantz et al. 2015; Pham et al. 2019). As an alternative to gene replacement, "suppression drives" aim to reduce the size of a pest population (e.g. Hammond et al. 2016; Kyrou et al. 2018). Gene drives that can cause very strong suppression may even be able to completely eradicate a population, and are here subcategorized as "eradication drives". As the level of suppression caused by a given gene drive can vary based on the ecological context of a particular population, the subcategorization of certain suppression drives as eradication drives is inexact. Similarly, the separation of gene drives into replacement or suppression categories is not absolute; some replacement drives could also cause some amount of population suppression. Even with these caveats, there is heuristic value in using these categories when discussing how different ecological factors affect the dynamics of gene drives.

Gene drives can also be categorized as threshold drives and non-threshold drives based on their frequency-dependent dynamics (Champer et al. 2016; Leftwich et al. 2018). Non-threshold gene drives exhibit a selective advantage over wildtype alleles irrespective of their frequency, and can spread even when they are initially very rare. In contrast, threshold drives suffer a selective disadvantage unless they are present above a certain threshold frequency (i.e. they exhibit bistability). Successful population alteration with threshold drives therefore requires them to be released in large enough numbers to exceed this "threshold release frequency" or "release threshold". Such release thresholds can result from fitness costs incurred by individuals due to carrying the drive alleles or can also be an intrinsic consequence of the drive mechanism. For instance synthetic Medea constructs and some CRISPR-based toxin-antidote drives, exhibit no threshold, unless the drive constructs impose a fitness cost on the carriers (even on drive-homozygous individuals) that is independent of the drive mechanism (Champer et al. 2020a,b; Oberhofer et al. 2019b; Ward et al. 2011). Even CRISPR-based homing drives can exhibit a release threshold under a limited set of conditions (Alphey & Bonsall 2014; Deredec et al. 2011; Tanaka et al. 2017), but it is unclear if such conditions can be met in natural environments. In contrast, underdominance mechanisms lead to a release threshold that has an intrinsic minimum level even if the drive construct bears no fitness cost (Sinkins & Gould 2006). The intrinsic minimum values for the release thresholds of different underdominance-based gene drives can differ widely based on the genetic structure of the drives (Altrock et al. 2010; Champer et al. 2019a, 2020a,b; Davis et al. 2001; Dhole et al. 2018; Leftwich et al. 2018; Marshall & Hay 2012a; Oberhofer et al. 2019a; Ward et al. 2011). Irrespective of whether a gene drive exhibits a non-zero intrinsic minimum release threshold, increasing the fitness costs suffered by gene drive carriers generally substantially increases the actual release threshold of a drive (Figure 1; e.g. Alphey & Bonsall 2014; Altrock et al. 2010; Champer et al. 2020a; Dhole et al. 2018; Ward et al. 2011; but see Dhole et al. 2019). Consequently, it can be difficult to categorize specific gene drives as non-threshold drives vs threshold drives, because environment-dependent changes in fitness can cause some gene drives to exhibit a threshold in some environments and not in others. But even a context-specific categorization can be useful for discussing the behavior of different gene drives.

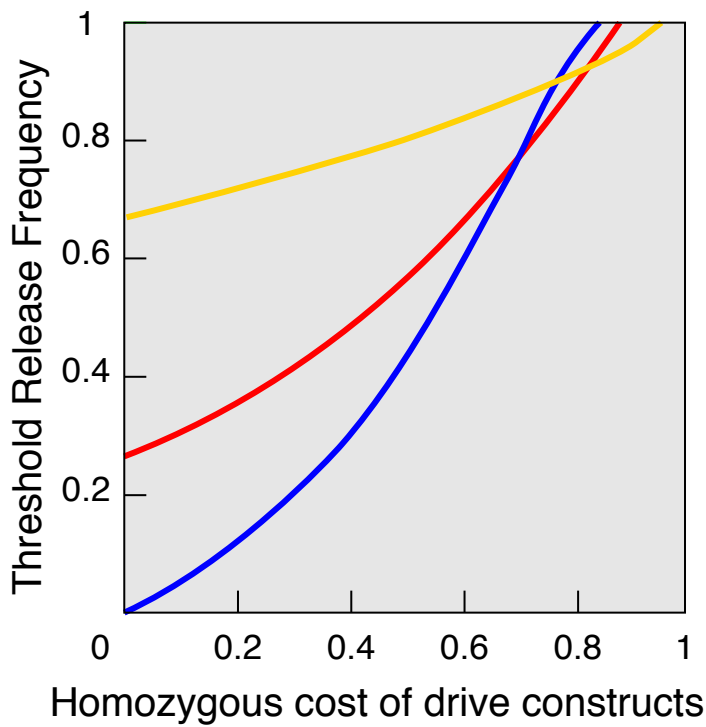

**Figure 1:** The threshold release frequency of threshold gene drives increases with the fitness cost imposed by the drive constructs. Shown here are the release thresholds for three gene drives – Medea (blue), 2-locus Engineered Underdominance (red), 1-locus Engineered Underdominance (yellow) – as they vary with the fitness cost incurred by individuals homozygous for the drive constructs.

In addition to requiring a larger release effort to exceed the threshold frequency during a release, the dynamics of threshold drives are qualitatively different from those of non-threshold drives, especially when spatial structure is considered (Barton 1979a). Moreover, threshold drives may also be less likely to invade non-target populations relative to non-threshold drives, because a larger cohort of migrants would be needed for surpassing the release threshold in a new population (Gould 2008). When confinement of a gene drive to specific populations is a concern, special caution must be exercised with efforts to use a gene drive that has a release threshold solely due to environmentally-dependent fitness effects, which may vary widely with time and between different environments (Backus & Delborne 2019).

A few gene drive approaches have been proposed that are temporally "self-limiting" (e.g. Killer-Rescue, Daisy-chain; Gould et al. 2008; Noble et al. 2019). These drives are designed to initially spread before removing themselves from the population eventually, and are not intended for permanent population alteration. Such drives may still be quite useful in some scenarios. For instance if a pathogen requires a critical mass of vectors for persistence, a temporary suppression of the vector population below such critical mass may be sufficient to eliminate the pathogen. The long-term dynamics of these drives, especially across space, are significantly different from self-sustaining drives. We focus this review only on self-sustaining gene drives.

In addition to gene drives, the cytoplasmic endosymbiont *Wolbachia* has been proposed as a tool for altering populations of certain insect disease vectors (Brownstein et al. 2003; Rasgon et al. 2003). These vertically transmitted endosymbionts can induce cytoplasmic incompatibility (CI) between gametes of infected males and uninfected females. While not a gene drive, CI-inducing *Wolbachia* can spread in an insect population with dynamics that are qualitatively identical to those of many threshold drives, and can reduce the transmission of certain insect-borne pathogens (Barton & Turelli 2011; Brownstein et al. 2003; Rasgon et al. 2003). Indeed, some of the most relevant recent work on the spatial spread of threshold drives is rooted in understanding the spread of *Wolbachia* (e.g. Barton & Turelli 2011).

Over the last decade, many aspects of gene drive technology have been reviewed, including the mechanisms of different gene drives (Burt 2014; Champer et al. 2016), biosafety aspects of gene drives (Marshall & Akbari 2018), diverse gene drive applications (Esvelt et al. 2014; Flores & O'Neill L. 2018; Gould 2008; Moro et al. 2018; Rode et al. 2019), the history of gene drive research (Macias et al. 2017), gene drive design considerations (Akbari et al. 2019), and the current status of development of certain drives (Leftwich et al. 2018; Raban et al. 2020). A recent review by Hay et al. (2020 *in press*) discusses progress towards new gene drive mechanisms that are being considered for achieving spatially restricted alteration, the long-term evolutionary stability of gene drives after release, and the hurdles that need to be navigated. Godfray et al. (2017) published a broad review on the use and dynamics of different endonuclease-based drives, addressing a number of factors that can affect drive spread in natural populations. David et al. (2013) have discussed the down-stream effects that gene drives could have on the ecology and evolution of altered populations. A number of committee and workshop reports have also focused on regulatory and practical considerations for gene drive applications (Delborne et al. 2018; Giese et al. 2019; James et al. 2018; National Academies of Sciences and Medicine 2016). In this review, we focus on key ecological factors that will play a pivotal role in the application of gene drives to natural populations. We specifically address topics that have not recently been reviewed in detail. These are 1) the effects of different forms of density-dependent population dynamics, 2) spatial structure within and between populations, and 3) mating behavior and sexual selection. We place special emphasis on spatial structure for which there is a great deal of recent and older relevant literature that has not been reviewed in the context of diverse gene drives.

**DENSITY-DEPENDENT POPULATION DYNAMICS:**

Population suppression gene drives can use two approaches for reducing population density (see reviews by Burt 2014; Champer et al. 2016; Sinkins & Gould 2006). The drive construct can be designed to disrupt an essential gene or include a harmful cargo gene so as to impose a direct fitness cost on individuals in the form of reduction in viability or fecundity (e.g. Hammond et al. 2018; Kyrou et al. 2018). Alternatively, but not mutually exclusively, the drive could be used to cause an extreme bias in the sex ratio of the offspring produced in the population (e.g. Galizi et al. 2014, 2016; Leitschuh et al. 2018; Prowse et al. 2019). A strongly female-biased sex ratio may reduce the number of progeny produced if many females are unable to find a mate and reproduce, while a male-bias reduces the number of females, directly reducing the number of offspring that can be produced. Both approaches, fertility- or viability-reduction and sex-ratio bias, are intended to impose a genetic load on the population, which is a measure of the extent to which the average fitness (reproductive capability) of a genetically altered population is reduced compared to that of a wild-type population, and which ranges between 0 (no reduction in population fitness) and 1 (complete reduction).

Different types of gene drives differ in their ability to impose genetic load on a population (Burt 2003; Champer et al. 2016; Deredec et al. 2008; Dhole et al. 2018; Khamis et al. 2018). Gene drives with high intrinsic minimum thresholds due to the drive mechanism (in this review: gene drives with a release threshold >0.5) require a large release effort even when individuals don't bear a fitness cost, and they may be unable to spread at all when they significantly reduce individual fitness. Gene drives with high intrinsic release thresholds are therefore thought to have limited potential for population suppression in most scenarios, but may find use for suppression in specific contexts (Champer et al. 2016; Dhole et al. 2018; but see Akbari et al. 2013; Khamis et al. 2018 for examples of temporary or specific-case population suppression). Gene drives with potential for having low or no release thresholds are generally able to spread even when they lower fitness of individuals, so they can impose a higher maximum equilibrium genetic load (e.g. Dhole et al. 2018, but see Champer et al. 2019a, 2020a; Dhole et al. 2019), making low (or zero) threshold drives especially powerful for population suppression (Alphey 2014; Burt 2003; Champer et al. 2016; Lambert et al. 2018; Leftwich et al. 2018; Sinkins & Gould 2006). While it is clear that a higher genetic load would tend to cause a stronger population suppression, exactly how strongly a given amount of genetic load can suppress a population is a far more complex question.

How a given amount of genetic load translates to population suppression critically depends on, among other factors, the extent and form of density-dependent dynamics in the population (e.g. Alphey & Bonsall 2014; Deredec et al. 2011). When intra-specific competition for resources is strong, reduction in population density is usually accompanied by increasing growth rates as the per capita resource availability increases (May 1973). This increase in growth rate with decreasing density can partially compensate for the suppressing effect of the genetic load.

The exact nature of this density-dependent response (e.g. strength and shape of the response) plays a critical role in determining how the equilibrium size of a population would change after imposition of a genetic load through a gene drive. Imagine a population where growth is restricted only by the level of resource competition so that the per capita growth rate is highest at an extremely low density and decreases to zero at the equilibrium size of the population (Figure 2a). All three curves in this figure exhibit this property but differ in the rate at which the per capita growth rate changes at a given density. If a population exhibits simple logistic density-dependent dynamics, so that the per capita growth rate changes linearly with density (Figure 2a curve ii), the size at equilibrium of that population would decrease linearly with the genetic load imposed by a gene drive at equilibrium (Figure 2b curve ii). Alternatively, in a population where the per capita growth rate decreases only slightly with increasing numbers at low densities, but very strongly as density increases to the equilibrium value (Figure 2a, curve i), a small reduction in density from the equilibrium value would cause very strong compensation in the growth rate (May 1973). In a population that exhibits such density-dependence, a much higher genetic load would be needed to achieve a similar level of suppression when compared to the logistic dynamics scenario (Figure 2b curve i). Conversely, if the per capita growth rate responds strongly to competition at low densities and is affected less by further increase in density (Figure 2a curve iii), even a small amount of genetic load may be sufficient to achieve a large amount of suppression (Figure 2b curve iii). This is because there is little compensation in the per capita growth rate as the population is reduced from its equilibrium (Figure 2a curve iii). For example, a model simulating a hypothetical mosquito population that exhibits density-dependence similar to curve iii was used to predict that an engineered underdominance drive imposing an equilibrium genetic load around 20-30% can cause close to 80% suppression of the mosquito population (Figure 3 in Khamis et al. 2018), resulting in a 50% reduction in disease transmission (Figure 4 in Khamis et al. 2018). In a population that exhibits logistic density-dependent dynamics, a similar drive (when released above its release threshold) would cause only around 20-30% suppression at equilibrium (e.g. Figure 2b curve ii). Factors that influence what form of density-dependent dynamics are exhibited by a given species (or population) are complex (Clutton-Brock et al. 1997; Herrando-Pérez et al. 2012; Lindström & Kokko 2002; Sinclair 2003). Understanding the density-dependent dynamics in the target population will be critical not only for accurate forecasting of the effect of a drive, but also for choosing and designing the gene drive that can achieve the target level of suppression.

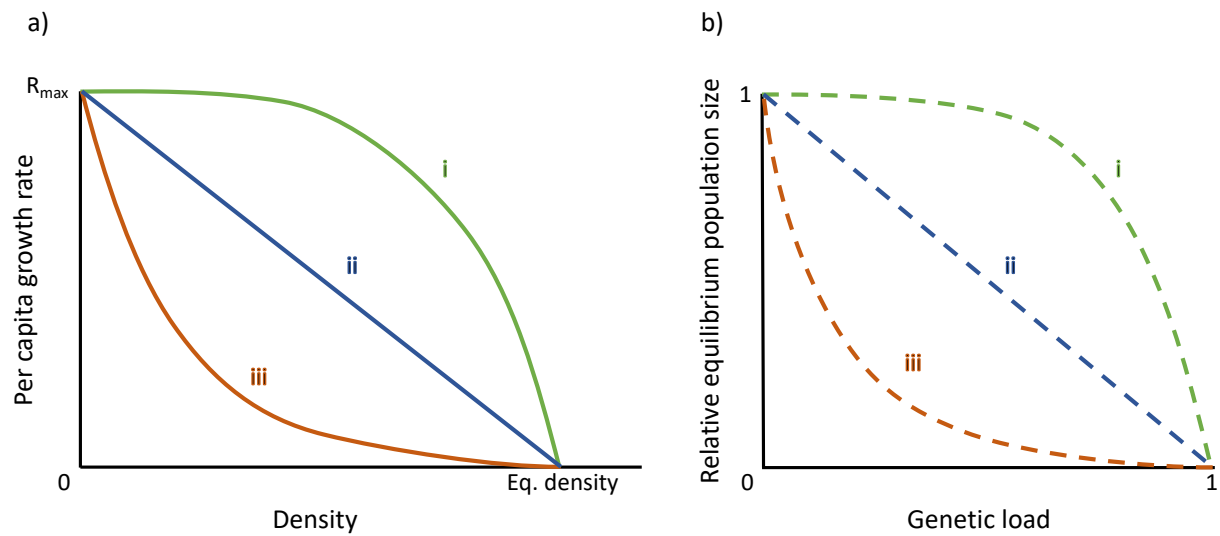

**Figure 2:** A highly simplified depiction of three forms of density-dependent dynamics is shown (a), along with corresponding curves for the relative equilibrium population size reached with a given amount of genetic load (b). $R_{max}$ denotes the maximum per capita growth rate at a low density.

While we have focused here on the functional types of density-dependent dynamics and their effect on suppression, other aspects of density-dependence can influence gene drive dynamics. A recent review by Godfray et al. (2017) discusses some of these in great detail, allowing us to only cover them briefly below. If density-dependent competition slows down development (e.g. Gimnig et al. 2002; Walsh et al. 2012), it can increase the generation time in a population and slow the spread of population replacement strategies (Hancock et al. 2016a). Strong Allee effects (Stephens et al. 1999), where a population exhibits negative per capita growth rate at very low densities, would also play an important role in population suppression. Strong Allee effects can aid in population eradication with ge drives by contributing to the reduction of replacement rate below 1 (Wilkins et al. 2018). Such Allee effects may also prevent subsequent recolonization of the emptied habitat by wild-type individuals. Another important consideration is the timing of gene drive effects. Suppression drives can potentially be designed to reduce fitness at different time points during an individual's development. If this fitness reduction occurs at a developmental stage before density-dependent competition occurs, the compensating increase in fitness of the remaining individuals can counter some of the drive's suppression effect. Timing fitness reduction to occur after the life stage where density-dependent competition is strongest (e.g. after larval stage in many insects) can produce greater suppression, because this also allows competition to reduce the population's reproductive output (Alphey & Bonsall 2014; Edgington & Alphey 2018; Yakob & Bonsall 2009).

Different species can certainly exhibit different forms of density-dependent dynamics (Bellows 1981; May 1973). But the relationship between per capita growth rate and density can vary widely even between populations of a single species or within a population with season, local genetic makeup, presence of other species et cetera (e.g. Rajagopalan et al. 1977; Walsh et al. 2013). Even different localities within a population could exhibit differences in density-dependence in heterogeneous habitats.

Unfortunately, density-dependence is poorly understood for natural populations of most pest species. Some empirical data are available for certain model organisms, some important disease vectors, and a few species of mammals (Bellows 1981; Chambers et al. 1999; Díaz et al. 2010; Gimnig et al. 2002; Hancock et al. 2016b; Lord 1998; Marlow et al. 2016; Muriu et al.

2013; Twigg & Kent Williams 1999; Walsh et al. 2012, 2013, 2011; Yoshioka et al. 2012). These data have been valuable for designing and parameterizing theoretical studies, but even in these taxa we know little about population- or environment-specific variation in density-dependent dynamics. Theoretical studies of gene drives have made different assumptions about density-dependent dynamics based on simplicity or on the biology of the specific taxon addressed. But different assumptions have been made by different models even when describing the same taxon (e.g. Akbari et al. 2013; Champer et al. 2019b; Eckhoff et al. 2017; Khamis et al. 2018; North et al. 2013, 2019; Sánchez C M. et al. 2019). Given the critical role played by density-dependence in determining the effect of gene drives on populations, we reiterate the call made by others before us for more empirical research on this topic (Beaghton et al. 2016; Godfray et al. 2017; Legros et al. 2009; Moro et al. 2018; Walsh et al. 2011).

We recognize the difficulty in gathering data that can allow determination of the functional form of density dependence. While a number examples of each form of dynamics have been suggested (e.g. Bull & Bonsall 2008; Sibly et al. 2005), much debate even exists about what form of density-dependent dynamics should be inferred from the same data (see Doncaster 2006; Getz & Lloyd-Smith 2006; Peacock & Garshelis 2006; Ross 2006). Factors such as stochasticity and seasonality in resource availability can also influence the apparent functional form exhibited by a population (Bull & Bonsall 2008; Lande et al. 1997; Lindström & Kokko 2002; Sæther 1997). In many cases, it may not be possible to determine the exact functional form of such dynamics with confidence. In these scenarios, mathematical models intended for forecasting population dynamics and for risk assessment would need to incorporate such uncertainties in the analyses.

**SPATIAL STRUCTURE WITHIN AND BETWEEN POPULATIONS:**

Few natural populations, if any, exist as a single well-mixed (i.e. panmictic and ecologically interacting) collection of individuals living across a uniform space. Most species occur in multiple, partially isolated populations in a heterogeneous landscape. Even within populations with a generally uniform distribution of individuals over a homogeneous landscape (continuous populations), individuals are more likely to interact with others that are in closer proximity. Therefore, in most populations, individual fitness is determined more by properties such as local density or local genotypic frequencies, rather than by the population-level averages for such quantities. Moreover, the fate of a gene drive released into a spatially structured natural population, depends disproportionately on the fitness of the individuals near the boundary of the release area (Barton 1979a; Barton & Turelli 2011). Whether the released transgenic individuals can spread the gene drive outward from the release area or are inundated and removed by incoming wildtypes is determined largely by the interactions in the zone where the cohorts of the wildtypes and transgenics meet. These details are lost in models that assume a well-mixed population. Many properties of a drive that affect these fine-scale, within-population dynamics are also important in determining how, and if, the drive spreads across a network of populations connected by migration. Understanding these spatial dynamics is crucial not only for optimal drive design and deployment, but also for understanding the potential risks of unintended spread.

A great deal of theoretical work has addressed how Mendelian and selfish genetic elements behave in spatially explicit landscapes. These studies range from simple 2-deme simulations through analytical diffusion models to biologically complex and highly detailed simulations of real geographical areas. Some of the theoretical underpinnings of the spatial dynamics of gene drives were revealed by the work of evolutionary biologists studying the spread of naturally occurring genetic elements (e.g Barton 1979a,b; Barton & Hewitt 1985, 1989; Lande 1985; Piálek & Barton 1997; Rouhani & Barton 1987). A number of recent studies have extended this body of work to *Wolbachia* and synthetic gene drives and have begun to incorporate more biological complexity (Barton & Turelli 2011; Beaghton et al. 2016, 2017b,a; Bull et al. 2019a,b; Champer et al. 2019b, 2020c; Eckhoff et al. 2017; Girardin et al. 2019; Hancock & Godfray 2012; Hancock et al. 2019; Huang et al. 2011; Legros et al. 2013; North et al. 2013, 2019; Tanaka et al. 2017). Below, we first discuss the work describing within-population spatial dynamics, and then address the effects of spatial structure on gene drive spread across a network of populations.

**Within-population dynamics:**

Consider the release of a large number of gene drive-carrying individuals in a certain area within an otherwise wild-type population. In real populations that are not perfectly mixed, spatial heterogeneity arises in a number of ways. First, the release of a large number of transgenics causes an increase in the local density (even if temporary), in turn affecting related factors like density-dependent dispersal or competition within that area (e.g. Backus & Gross 2016). Second, the genetic composition of potential mates available to individuals near the edge of such a release area or of an area where the gene drive has already been established is very different than that available to individuals at the center of or completely outside that area. The gene drive would spread out if the frequency of drive alleles increases at the forefront of the boundary of the release area or of an area where the drive is already established. If instead, the wildtype frequency keeps increasing at the boundary, the boundary would be pushed inward, causing drive failure. The change in allelic frequencies at the boundary is determined by two factors – 1) the frequency-dependent relative fitnesses of the alleles (transgenic and wild-type) at the boundary, and 2) the addition of alleles through dispersal from either side of the boundary. Fisher (1937) used a reaction-diffusion framework to describe these dynamics at the boundary for beneficial mutations as a wave of genotypic frequencies. Barton (1979a) extended this work to describe the spread of natural genetic elements that exhibit a threshold behavior similar to threshold gene drives. While the reaction-diffusion framework requires many simplifying assumptions, it has been very useful for understanding the spatial behavior of genetic elements without a frequency threshold for their fitness relative to wildtypes ("Fisherian waves") and those with a frequency threshold ("Bartonian waves"). The spatial dynamics of these two types of genetic elements are significantly and qualitatively different. Below we first highlight how the Fisherian and Bartonian dynamics differ, and then separately consider the dynamics of threshold and non-threshold drives under the influence of different ecological factors.

*Fisherian vs Bartonian waves:* At the forefront of the boundary of a release area, the "leading edge" of the wave (Figure 3a), drive alleles are by definition going to be rare initially. Non-threshold drive alleles have an evolutionary fitness advantage over wild-type alleles irrespective of their frequency and can increase in frequency even at the leading edge of the wave. Thus, regardless of the fate of the drive carrying individuals behind the leading edge of the wave, just the evolutionary fitness advantage at the forefront of the boundary can pull the wave of a non-threshold drive forward (Figure 3b). In contrast to these Fisherian wave or "pulled wave" (Stokes

1976) dynamics of non-threshold drives, threshold drives cannot increase in frequency through an evolutionary fitness advantage at the leading edge where they are rare. Threshold drives therefore critically rely on dispersal (spillover) of drive carrying individuals from areas where they are at high frequency into the leading edge for the wave to move forward (Figure 3c). Movement of this Bartonian or "pushed" wave-front of threshold drives is intimately tied to the fate and dispersal of drive-carrying individuals behind the leading edge (Barton 1979a; Beaghton et al. 2016; Godfray et al. 2017; Stokes 1976).

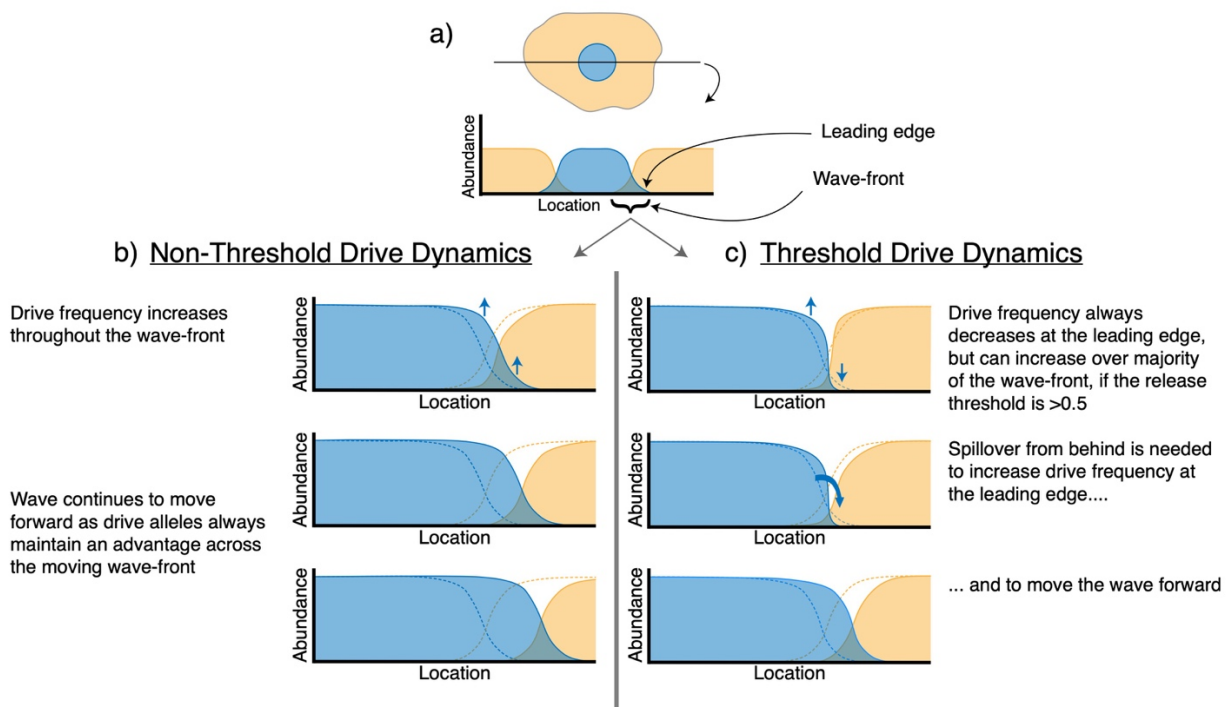

**Figure 3:** Fisherian versus Bartonian wave spread. Dynamics are shown after release of a gene drive (blue) within a wild-type (orange) population. Allelic frequencies at a cross-section of the landscape are shown after local drive establishment at the release area. Dynamics at a wave-front (region shown by curly bracket) are shown for two types of gene drive. Original wave positions shown with dashed curves.

***Non-threshold drives:*** Gene drives that exhibit Fisherian dynamics can establish a spreading wave even in absence of any additional transgenics dispersing in from behind the wave (Figure 4a; e.g. Beaghton et al. 2016; Tanaka et al. 2017). While any spreading wave requires forward dispersal *from* the leading edge, the lack of need of dispersal *into* the leading edge from behind has important implications for population suppression or eradication with non-threshold drives. A non-threshold eradication drive can spread across a continuously distributed but non-panmictic population even if it leaves empty space in the wake of the wave, similar to a spreading ring of forest fire (Figure 4c; Beaghton et al. 2016; Tanaka et al. 2017). This could result in eradication of an entire uniform, continuous population, assuming there is no long-range dispersal that could allow wildtypes to recolonize the empty space. While reaction-diffusion models assume only very short-range dispersal, stochastic, individual-based spatially explicit models of eradication drives show that if wildtypes can disperse to occupy the empty space created by an eradication drive, perpetual local cycles of drive invasion, eradication and wild-type recolonization can occur in continuous but non-panmictic populations (Champer et al. 2019b) or in structured metapopulations (Bull et al. 2019b; Eckhoff et al. 2017; North et al. 2019). Similar cyclical spatio-temporal dynamics are also predicted for certain reversal drives released to eliminate a previously established homing gene drive, between the alternative states of local occupation by wildtypes, homing drive, and reversal drive in spatial and non-spatial contexts (Girardin et al. 2019; Vella et al. 2017). Such dynamics can interfere with complete eradication of the population. However, if strong Allee effects prevent or slow down the establishment of a viable population in the eradicated area by a small number of dispersing wildtypes, complete eradication may still be feasible. Repeated introductions of the gene drive may also help to reduce such perpetual dynamics (Eckhoff et al. 2017; North et al. 2019). Gene drives that suppress a population, but do not cause eradication, may avoid such cycles, because low but persistent presence of individuals carrying such a drive in the suppressed areas could resist recolonization by a small number of incoming wildtypes. If the epidemiological, conservation or agricultural purpose behind population alteration does not require eradication, a less powerful suppression drive could offer an efficient solution.

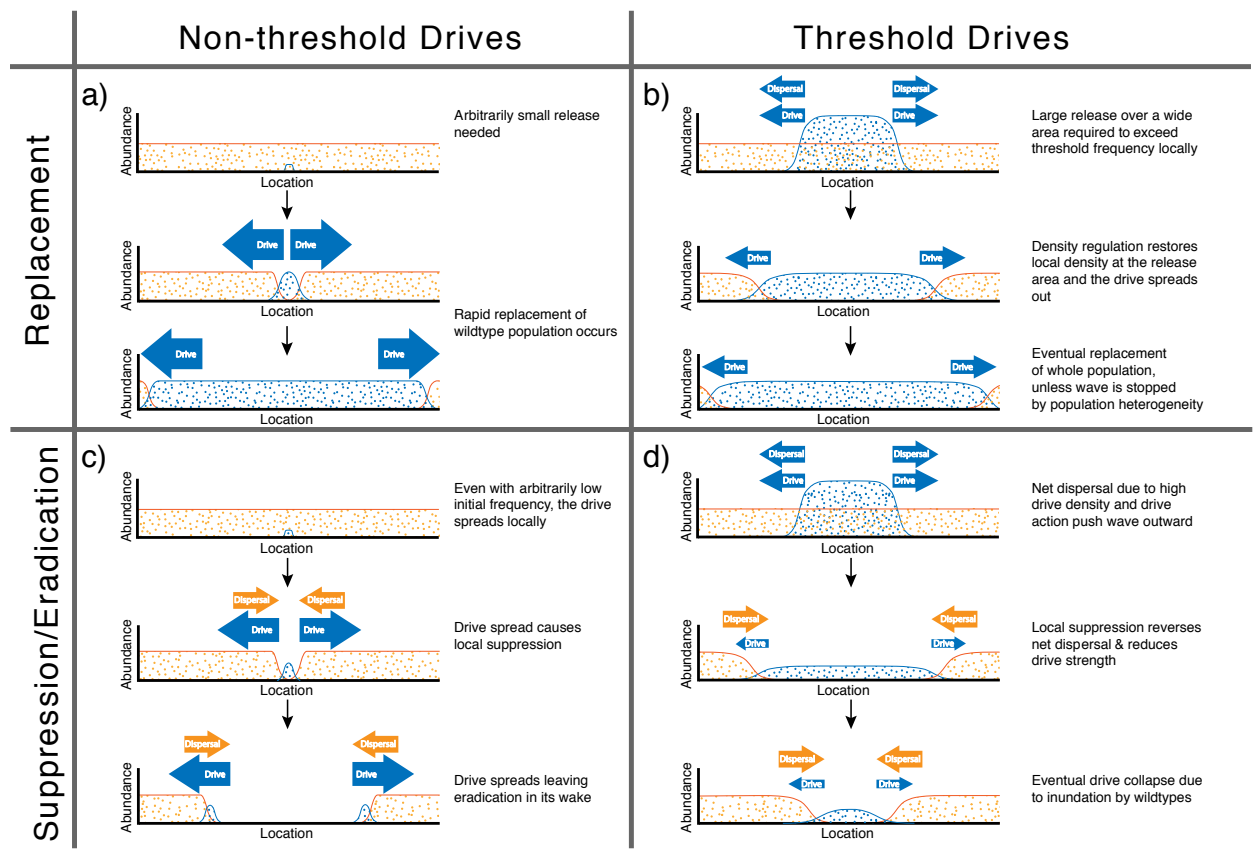

**Figure 4:** The dynamics of replacement (top row) and suppression or eradication (bottom row) are shown for non-threshold (left) and threshold drives (right). Within each panel, black arrows indicate progression of time. Horizontal arrows show direction and relative strengths (indicated by arrow size) of factors that favor, in particular contexts, spread of the drive (blue arrows) or of the wild-type alleles (orange arrows).

In contrast to suppression drives, non-threshold, replacement drives are expected to be pulled forward with no impact of wildtypes dispersing into areas that have already been occupied by individuals with the gene drive and altered phenotype. Of course, in more realistic scenarios with spatial heterogeneity, seasonality, evolution of resistance alleles, molecular breakdown of gene drive components and other complexities, even non-threshold replacement drives can fail to spread or achieve the desired outcome (e.g. Beaghton et al. 2017b; Bull 2017; Bull et al. 2019a,b; Girardin et al. 2019).

***Threshold drives:*** As mentioned above in 'Fisherian vs Bartonian waves', an increase in the frequency of a threshold drive at the leading edge of the boundary of the release area depends critically on addition of transgenes from behind the leading edge through dispersal (Figure 3c; Barton 1979a). The movement of such pushed waves over continuous space is therefore affected by all the factors that influence net dispersal across the wave-front, in addition to the relative evolutionary fitness of the drive and wild-type alleles (Barton 1979a; Barton & Hewitt 1985, 1989; Barton & Rouhani 1991; Barton & Turelli 2011; Michalakis & Olivieri 1993; Piálek & Barton 1997; Rouhani & Barton 1987).

The threshold release frequency of a gene drive determines the net selective advantage (or disadvantage) experienced by drive alleles over the width of the wave-front in continuous populations. Assuming equal local densities in the area behind a wave (where the drive has established) and the area occupied by wildtypes in front of the wave, and no differences in dispersal between transgenics and wildtypes, a gene drive with a release threshold much greater than 0.5 will always face a selective disadvantage across a wave-front. Reaction-diffusion models therefore lead to the conclusion that, all else being equal, gene drives with a release threshold much greater than 0.5 cannot spread out, even when the initial drive frequency in the release area is far greater than the release threshold (Barton & Turelli 2011). On the other hand, a wave of a gene drive with a release threshold lower than 0.5 (e.g. Champer et al. 2020b; Dhole et al. 2019; Oberhofer et al. 2019a; Ward et al. 2011) may be able to move forward (outward from the release area) across a uniform landscape (Figure 4b; Barton & Turelli 2011). Later in the manuscript we discuss possible deviations from and caveats for this conclusion. This conclusion is also our motivation for using a release threshold of 0.5 to distinguish between "high-threshold" and "low-threshold" (see below) drives in this review. This terminology is used here solely for the purpose of the discussion in this review, and is not intended to contend with other definitions of these terms (James et al. 2018; National Academies of Sciences and Medicine 2016).

Even for gene drives with a release threshold <0.5 (here referred to as "low-threshold drives"), their frequency at the leading edge can increase only through dispersal of drive alleles from behind. Therefore, a reduction in the net dispersal of drive alleles into the leading edge can halt the spread of threshold drives (Barton 1979a; Barton & Turelli 2011; Piálek & Barton 1997). A difference in density of individuals between two regions can give rise to a bias in net dispersal of alleles from high- to low- density areas, simply because there are more individuals that can disperse from the high density region. Such density gradients can occur due to heterogeneity in the distribution of resources that determine local carrying capacity or due to the fitness effects of the gene drive itself. If a traveling Bartonian wave of a threshold drive encounters a density gradient, with density increasing in the direction of wave spread, the spillover of drive alleles into the leading edge of the wave from behind may get outnumbered by inundation of wildtypes from the more dense area in front of the wave (Barton & Turelli 2011). Conversely, a density gradient with density decreasing in the direction of gene drive spread can favor the spread of the drive;

for example, in a scenario where the drive release occurs in an area of high carrying capacity (e.g. Champer et al. 2018). A wave of a spreading threshold drive could also be halted by a zone of low density because of the unfavorable density gradient on the opposite side of the zone (Barton 1979a; Piálek & Barton 1997). Such effects of density gradients can have an especially strong influence on threshold drives that cause population suppression. Consider a threshold drive such that transgenic homozygotes have lower viability (or reproductive ability) relative to wild-type homozygotes. If this disadvantage, which is frequency-independent, results in a lower local density in the area where the drive is present (i.e. if the drive causes local population suppression), net dispersal would act against drive spread (Figure 4d; Barton 1979a; Barton & Turelli 2011). Such a dispersal disadvantage would certainly further reduce the ability of spatial spread for high-threshold drives, but if the density reduction due to the gene drive is strong enough, even low-threshold drives could fail to spread.

Theoretical studies that have examined the effects of density gradients on the spatial dynamics of threshold drives (or natural elements that exhibit bistablility) have largely focused on density gradients as a property of the environment, rather than resulting from fitness effects of different genotypes (e.g. Barton 1979a; Barton & Hewitt 1985; Barton & Rouhani 1991; Barton & Turelli 2011; Champer et al. 2020c). If genotypic fitness can affect local density, the form of density-dependent dynamics could have a strong influence on the formation of density gradients, and in turn, on the spatial dynamics of gene drives. For instance, in a population with density-dependence described by curve iii in Figure 2 (e.g. Sibly et al. 2005), even small fitness costs of the drive constructs can cause a large difference in the densities on either side of the wave, limiting the spread of the drive even if the threshold release frequency of the drive remains the same. We are not aware of any analytical diffusion model of Bartonian wave dynamics that allows a strong effect on local density of a spreading genetic element of interest (see results for a Fisherian wave in Beaghton et al. 2016). But some recent individual-based simulation studies suggest that genotype-mediated changes in local density play an important role in the spread of threshold drives (Champer et al. 2019b, 2020c; Huang et al. 2011). Theoretical studies that can incorporate different forms of density-dependent dynamics into models of spatial spread of gene drives would be valuable for understanding how threshold (and non-threshold) gene drives would behave in natural population of different species or in different environmental conditions. Such models will also be useful for shedding light on the potential applicability of threshold drives for suppression of spatially structured populations, which thus far has only been addressed in panmictic scenarios (e.g. Akbari et al. 2013; Dhole et al. 2018, 2019; Khamis et al. 2018).

Much like density gradients, an unfavorable gradient in dispersal or a barrier to dispersal can similarly halt the spread of a threshold drive (Barton 1979a). In the case of urban dwelling mosquitoes such as *Aedes aegypti*, the landscape between cities or towns is expected to be such a barrier. Even features such as roads and rivers may provide strong enough dispersal barriers to halt the spread of certain threshold drives. In general, environmental heterogeneity is likely to limit the spread of threshold drives across a landscape.

Another factor in many pest species that adds complexity to spatial dynamics of gene drives is a separation between the life stages that face density-dependent regulation and the life stages that disperse and reproduce. In a model of an eradication driving-Y gene drive, which exhibits Fisherian-wave dynamics, Beaghton et al. (2016) showed that the speed of wave spread for the drive decreases in proportion with increasing fraction of total lifetime spent in immobile juvenile stages. The spread of threshold drives, which relies critically on dispersal of transgenics from behind the leading edge of the wave, may be especially sensitive to a lag between density-regulation and dispersal. Spatially explicit models that can incorporate such developmental delays will be useful for understanding how rapidly threshold drives can spread through such stage-structured populations, such as those of disease-vectoring mosquitoes.

***Critical bubbles and release strategies for threshold drives:*** The release of a threshold drive certainly needs to be large enough to exceed the release threshold within the release area. But, how much should the threshold be exceeded by, and how large should the release area be? The answer, of course, is not straightforward. Barton and colleagues (Barton 1979a; Barton & Turelli 2011; Rouhani & Barton 1987) have derived expressions that describe the critical frequency distribution over space, termed the "critical bubble", that needs to be exceeded to establish a spreading wave for simple underdominance elements and for *Wolbachia* that can induce cytoplasmic incompatibility. While these expressions rely on some simplifying assumptions that may not be realistic for many scenarios of gene drive releases, they have high heuristic value. They show that for threshold drives, the threshold release frequency needs to be exceeded over a release area with a minimum critical radius. Initial establishment of the drive over a large area helps the wave spread in two ways. First, release over a larger area reduces the dilution effect of dispersal; a large area is less likely to be inundated by wildtypes from the surroundings than a small one. Second, the geography of the boundary of the release area matters. A strong curvature of the boundary results in a bias in the net dispersal from outside the curve to inside (Barton

1979a; Champer et al. 2020c). Increasing the release area can reduce the curvature of the boundary, aiding the spread of the drive to some extent.

The analysis of a reaction-diffusion model by Barton and Turelli (2011) suggests three interesting things about the critical radius. First, the critical radius increases very slowly with increasing release threshold of a gene drive until the threshold begins to exceed ~0.3, after which the critical radius increases very rapidly. Second, while the initial release frequency needs to exceed the threshold release frequency, the magnitude of this excess has only a small effect on the critical radius. Third, the absolute critical radius certainly increases with the dispersal rate of individuals. Similar expressions are not yet available for other types of gene drives that exert strong selection (e.g. suppression drives or drives that cause strong reduction in heterozygote fitness) or have more complex genetic structure than simple one-locus underdominance. One caveat to these results regarding the critical radius is that this analysis does not account for the increase in local density, even if temporary, that must come with a gene drive release. Increasing the initial release frequency of a gene drive, for instance from 0.5 (1:1 drive:wildtype) to 0.8 (4:1 drive:wildtype), can correspond to a significant increase in the density gradient at the boundary of the release area. Even if the gradient remains short-lived due to local density regulation, it may have a significant impact on the critical radius required to get a spreading wave established. It remains to be seen how large an effect such density changes can have on the critical radius. For application to gene drive release in natural populations, estimating the critical bubble size would need much extended and more case-specific analyses that can account for factors such as sustained differences in density due to repeated releases of the drive, the density-dependent dynamics that can remove density gradients, heterogeneity in the carrying capacity across space and time et cetera (e.g. see Champer et al. 2018)

The discussion above is restricted to the analysis of a single release in one area. However, the spatial patterns of release can play a large role in establishment and spread of threshold drives. Huang et al. (2011) explored a model of an age-structured population in a 32x32 grid of discrete but highly connected patches. Individuals with a 2-locus engineered underdominance drive (intrinsic minimum release threshold ~0.27) were either released in a central core of patches (central release) or into an equivalent area of patches that were uniformly distributed over the entire grid (distributed patches). They found that total number of released transgenic individuals needed to establish the drive depended on many factors including age of released individuals, fitness costs, and migration rates. An interaction between dispersal rate and fitness differences between drive and wild-type individuals determined whether a smaller number of total released individuals were needed in the central releases or the distributed patch releases. Detailed simulation models of mosquito populations also suggest an important role of multiple releases distributed across the landscape (Eckhoff et al. 2017; Legros et al. 2013; North et al. 2019). These models demonstrate that while the general principles from simple reaction-diffusion models are upheld, biological properties of the target species will be critical in determining optimal gene drive design and release strategies.

Empirical data on these biological properties are not available for most of the species proposed as targets for gene drives. However, a number of recent empirical studies have focused on elucidating the spatial ecology of important disease vector species (e.g. Dao et al. 2014; Epopa et al. 2017; Goubert et al. 2016; Guagliardo et al. 2019; Huestis et al. 2019; Kotsakiozi et al. 2018;

Lehmann et al. 2017; Rašić et al. 2015, 2014). More studies on the ecology of pests are critically needed.

**Between-population spread:**

Migration rates are expected to impact the spread of both threshold and non-threshold gene drives between populations with finite size. However, the migration rates needed for any establishment of threshold drives in a new population are much more substantial (Dhole et al. 2018; Láruson J. & Reed 2016; Marshall & Hay 2012b; Noble et al. 2018). Modeling efforts on between-population spread have been mostly focused on threshold drives because the goal of threshold drive developers is often prevention of cross-population spread and models that can provide insights into the risk of a threshold drive spreading will be of special interest to regulatory authorities (Backus & Delborne 2019). There are similarities between factors that impact the spread of gene drives in single non-panmictic populations and those that impact spread between discrete populations. But, there are also qualitative differences in the dynamics because the speed of local population dynamics can be much faster than the speed of dispersal dynamics between discrete populations.

As a first approximation of inter-population spatial structure, a number of theoretical studies have used a simple scenario where two (or very few) large, well-mixed populations exchange migrants at some specified rate every generation (e.g. Akbari et al. 2013; Altrock et al. 2010; Dhole et al. 2018, 2019; Láruson J. & Reed 2016; Marshall & Hay 2012b; Noble et al. 2019; Sudweeks et al. 2019). While highly simplified, this scenario provides a very convenient method for studying basic properties of gene drive spread between populations. Such models are not intended to forecast the exact dynamics of a gene drive in natural conditions, but rather are used as tools to describe basic gene drive properties, and importantly, give comparative measures of the tendencies of different gene drives to spread to new populations through migration (e.g. Dhole et al. 2018; Marshall & Hay 2012b). They can also help in determining how different properties of the gene drive or of the two populations can affect the drive's ability to spread across population boundaries (e.g. relative fitness of transgenics, the level of haploinsufficiency of target genes, level of resistance in neighboring populations etc.; Dhole et al. 2019; Noble et al. 2019; Oberhofer et al. 2019; Sudweeks et al. 2019).

As mentioned above, the within-population spread of non-threshold gene drives is largely independent of the dynamics behind the wave-front, making even complete eradication feasible (with caveats regarding recolonization by wildtypes; Beaghton et al. 2016; Champer et al. 2019; Okamoto et al. 2013). Spread across different populations, however, can be difficult for even non-threshold eradication drives, because if the source population is eradicated very rapidly, the drive may not have a chance to spread to a new population (Eckhoff et al. 2017; North et al. 2013). The fate of an eradication drive, and also of strong suppression drives, is highly dependent on the relative rates of local eradication, dispersal across populations and recolonization by wildtypes (Eckhoff et al. 2017; Lande 1985; North et al. 2013, 2019). As in the case of a population uniformly distributed over a continuous landscape (Champer et al. 2019b; Girardin et al. 2019), for a range of realistic dispersal rates, perpetual cycles of drive invasion, local extinction and wildtype recolonization can occur across a large network of connected populations (Eckhoff et al. 2017; North et al. 2013, 2019). If individual populations are sufficiently small to make stochastic local extinctions common, the spread of replacement drives across multiple populations also becomes less certain. Moreover, if populations experience extreme seasonal fluctuations, stochastic local extinctions may become more likely during seasonal lows (Eckhoff et al. 2017). Seasonal changes in dispersal may also influence how gene drives spread between populations (North et al. 2013, 2019). Temporal variation in dispersal and fitness has not yet received substantial formal investigation with regards to gene drives, but it will certainly play an important role in spread between populations.

Threshold drives need a much larger cohort of transgenics to establish a drive in a new population, so much higher dispersal rates would be required them to invade new populations than needed for non-threshold drives. Very high dispersal rates can also cause the source population to be inundated by incoming wildtypes and cause the drive to be lost from all populations (e.g. Dhole et al. 2018; Láruson J. & Reed 2016). Threshold drives are therefore unlikely to be suitable for alteration (even replacement) of multiple populations, unless large releases are carried out in each targeted population (but see Dhole et al. 2018; Láruson J. & Reed 2016; Marshall & Hay 2012b). If threshold drives were to be used for eradication or suppression, they would face an even stronger challenge when spreading across multiple populations, assuming they can spread within a population in the first place. This theoretical inability of threshold drives to spread across populations, or in other words, the ability to remain localized to a population is attractive when spread across multiple populations is undesirable (Buchman et al. 2018b; Curtis & Robinson 1971; Davis et al. 2001; Dhole et al. 2018; Magori & Gould 2006; Marshall 2010; Marshall & Hay 2012b; National Academies of Sciences and Medicine 2016).

The theoretical studies discussed above that analyze the likelihood of gene drive spread across multiple populations assume panmictic populations or patches that exchange migrants (e.g. Dhole et al. 2018, 2019; Huang et al. 2011; Láruson J. & Reed 2016; Marshall & Hay 2012b; Noble et al. 2018, 2019). In such simple models, drive-carrying individuals migrating into a new population are expected to interact largely with wildtypes that have a much higher overall frequency. In real scenarios, however, migrating individuals are more likely to arrive in a certain area of the new population. For example, transgenics migrating via ships would arrive at a port, or populations connected by narrow corridors (e.g. highways between towns) would arrive at the end of the corridor. A recent study by Champer et al. (Champer et al. 2020c) modeled a two-deme scenario with the populations within each deme spread over a continuous space. The study shows that when migrants are more likely to arrive in close proximity, the likelihood of a threshold drive invading the new population is much higher than expected from panmictic deme models. This is an important result and clearly calls for empirical research to assess spatial patterns of migration into new populations. Of course, real populations are not continuously distributed. Discrete barriers such as rivers, roads, human habitations, or even the tendency of individuals to be located near discretely dispersed resources like ponds can result in small patches of individuals that are panmictic within the patch and somewhat isolated from others. Gene drive spread between such heterogeneous populations might be more realistically simulated by population network models where each population is modeled as a closely connected group of panmictic demes (e.g. Huang et al. 2011; North & Godfray 2018; North et al. 2019; Sánchez C M. et al. 2019).

As discussed above, strong population structure is likely to make gene drive spread more difficult, especially for eradication drives, which may not be able to spread to all populations before going extinct (Eckhoff et al. 2017; North et al. 2019). One possible solution to reduce the likelihood of local extinction may be to repeatedly introduce the drive in multiple changing locations (e.g. North et al. 2019). Long-term and large-scale eradication of pest species will require more effort in spatially structured populations than estimated by panmictic models. On the other hand, spatial structure could make it easier to effect changes in local populations, when global spread is undesirable. Any genetic differentiation between highly spatially structured populations may also allow the use of population-specific molecular confinement mechanisms (see below).

*Localized population alteration:* Non-threshold drives may provide a mechanism for altering natural populations on a large scale, but they pose a challenge for spatial containment when only local population alteration is desired (National Academies of Sciences and Medicine 2016). As we have mentioned before, threshold drives could provide a solution in such cases, especially low-threshold drives. A number of threshold drive designs fit into the category of low-threshold drives, especially if the drive constructs themselves impose only a small cost on the carriers (e.g. Akbari et al. 2013, 2014; Champer et al. 2020a,b; Davis et al. 2001; Marshall & Hay 2011; Marshall et al. 2011; Oberhofer et al. 2019a; Ward et al. 2011). Such low-threshold drives may be able to achieve local population alteration without requiring very large releases in the target populations. However, if a gene drive construct imposes a significant cost on the carrier, this cost is likely to shift the release threshold to a high value (e.g. Dhole et al. 2018, 2019; Leftwich et al. 2018; Ward et al. 2011), which would raise the amount of release effort required and also increase the likelihood of inundation by incoming wild-type migrants. Therefore, while a number of mechanisms may be available for localized population replacement, localized population suppression remains a challenge.

A few gene drives that exhibit a release threshold solely due to fitness costs incurred by individuals carrying the gene drive have been proposed for localized population suppression (Champer et al. 2020a; James et al. 2018). If the fitness costs to the carriers remain stable across time and environment, such gene drives would remain spatially restricted. However, environment-dependent or evolved changes in fitness of drive-carriers after the release may unpredictably change or even remove the release threshold and thus the localization ability of such gene drives. As we stated before, special care would need to be taken when using such gene drives for spatially restricted use, for instance, to ensure that the fitness costs incurred by different genotypes remain within bounds that allow maintenance of desired release thresholds for the entire period that the gene drive remains active in the population. Gene drives that exhibit intrinsic minimum release thresholds may provide a form of added insurance against such changes in fitness. We recently proposed two designs for new gene drives that may provide a solution – Tethered Homing drive (TH) and a homing drive targeting locally fixed alleles (Dhole et al. 2019; Sudweeks et al. 2019).

The TH design is based on driving a fitness-reducing homing construct that is anchored to a threshold component (Dhole et al. 2019). The homing construct is essentially a fitness-reducing suppression HEG drive with its Cas endonuclease removed, disabling its ability to spread rapidly on its own. The threshold component is any threshold drive that can be linked with genes for germline-specific expression of a Cas endonuclease. Because of its release threshold the threshold component can spread only in the target population where it is released in sufficient numbers, thus establishing a localized population of individuals that express the Cas endonuclease in their germline. In absence of the homing construct, the Cas endonuclease has no activity, because it lacks any guide RNAs. The homing construct, with its Cas endonuclease removed, can spread through gene conversion only when it co-occurs with the threshold component. Any homing constructs that reach a non-target population cannot spread there in absence of the Cas-bearing threshold component and will be removed by natural selection due to their fitness costs. This dependency of the homing construct on a component that is able to remain localized can allow it to spread rapidly in the target pest population, but can prevent it from invading non-target populations. A major benefit of the TH design over other threshold drives is the separation of the threshold component from the high fitness costs required for population suppression, which come from the homing construct in a TH drive system. Thus, unlike other threshold drives the release threshold of the TH drive remains largely unaffected by the fitness costs imposed by the TH drive, allowing the possibility of strong localized suppression without the need for a very large release effort. The intrinsic minimum release threshold of the threshold component can provide the insurance against environment-dependent changes in fitness mentioned above. Another benefit of this system is that the threshold component provides a persistent, spatially restricted source of Cas endonuclease that can be used to spread additional homing components targeting other genes for stronger population suppression or population alteration. Since the threshold component could be any threshold drive that can be linked with Cas endonuclease genes, this system may provide additional flexibility for creating a drive in different species.

Taking a very different approach to local containment, one that is not based on a release threshold, Sudweeks et al. (2019) have proposed a method for localizing non-threshold CRISPR-based homing drives to islands where invasive pests are targeted for eradication. The method relies on the fact that on isolated islands, drift and localized selection are likely to cause fixation of alleles that are polymorphic on the mainland. If these locally fixed alleles are used as targets for a suppression homing drive, the drive is expected to rapidly spread on the island by targeting the locally fixed allele, but it will not permanently establish in mainland populations where homologs of the fixed island allele are present and cannot be targeted by the homing drive's guide RNAs. While this approach may work for islands, it is much less likely that two large mainland populations will have differential fixation of many targetable alleles, making the method less suitable for locally altering mainland populations.

Both these strategies have very specific requirements for achieving successful localized spread (Dhole et al. 2019; Sudweeks et al. 2019), but they provide a rare set of solutions for safer locally restricted, yet efficient population suppression (Hay et al 2020 in press). It is our hope that ongoing progress in genetic engineering techniques will facilitate creation of these systems in pest species important in epidemiological and conservation applications.

**SEX AND MATING:**

All gene drives depend on mating and fertilization between transgenic and wild-type individuals. Mate choice by wild-type females against drive-carrying males can strongly hamper the spread of gene drives (Huang et al. 2009, 2011; Khamis et al. 2018; Leftwich et al. 2016; Manser et al.

2015, 2017; Wedell 2013). While females of some species can discriminate against males that carry certain natural meiotic drivers (Wedell & Price 2015), it is not clear if females could quickly evolve to directly discriminate against synthetic gene drives that can spread rapidly through a population. However, a number of existing behaviors can create a mating bias. For instance, if adult females mate only during limited periods in their lifetimes, the number of wild-type females that are available to mate at a given time could be low. For instance, *Aedes* mosquitoes mate soon after eclosing and usually mate only once during their lifetimes in the field (Foster & Lea 1975; Williams & Berger 1980; Young & Downe 1982). Therefore, the only wild-type females available for mating with transgenic males would be females that eclose after the release. Even if adult females mate many times and don't have second-male sperm precedence, it must be remembered that the majority of individuals in an insect population are typically found in the immature stages and are therefore unavailable for mating. When calculating an appropriate release threshold, it is important not just to consider the number of individuals released, but also their reproductive value (Fisher 1930) and the reproductive value of the individuals with whom they mate (Huang et al. 2009; Legros et al. 2013). In a model of gene drive spread in an age-structured population, Huang et al. (2009) showed that low levels of polyandry can significantly hamper the spread of a gene drive, especially in cases of male-only releases. Even if only females instead of males are released there can be limitations to mate availability. This limitation can be partially mitigated by bi-sex releases (Huang et al. 2009; Khamis et al. 2018; Legros et al. 2013; Robert et al. 2013). In many species, however, the release of females may not be feasible due to the disease transmitting capabilities of females (e.g. in mosquitoes) or due to the potential increase in population growth (even temporary) due to the addition of females. Another example of behaviors that can reduce matings between drive and wild-type individuals is inbreeding, caused either by strong population structure or by active mate choice. Two models by Bull et al. (2017; 2019b) show that even low levels of inbreeding (or especially selfing in plants) can strongly hamper of the spread of gene drives.

In polyandrous species, where females mate with multiple males within a breeding cycle, sperm competition can hamper the reproductive ability of drive-carrying males even in the absence of mate choice (Lindholm et al. 2016; Wedell 2013). There is considerable evidence that certain sex-ratio distorting drives can reduce males' sperm-competitive ability in a number of taxa (Price & Wedell 2008). The t-haplotype in mice is a large cluster of genes that biases its transmission through heterozygous males by rendering sperm without the t-haplotype dysfunctional (Herrmann & Bauer 2012). In an experimentally evolving population of mice, Manser et al. (2017) showed that the spread of a t-haplotype is significantly hindered in populations with strong sperm-competition. A gene drive that links a sex-ratio distorter to the t-haplotype is being considered for eradication of invasive mice on islands (Godwin et al. 2019; Leitschuh et al. 2018). A sperm-competitive disadvantage could enforce a much larger release effort for successful alteration with such a gene drive.

Sex-ratio biasing drives that cause a scarcity of males can suppress a population if a large number of females are unable to find a mate. The level of polygyny, i.e. the mean number of mates per male, can severely affect the efficacy of such gene drives. For example, a Y-shredder drive that reduces the number of males would require a much stronger bias in the sex-ratio in a polygynous population than in a monogamous population to achieve the same level of suppression (Prowse et al. 2019). If the female-biased sex ratio is caused through reduction in male-forming sperm, most of the offspring are born as females, resulting in an increase in the absolute number of females. In populations where a small number of males can fertilize a large number of females, such sex-ratio bias could actually increase population size, because it increases the number of fertilized females (Prowse et al. 2019). Male-reducing gene drives that do not increase the absolute number of females (e.g. by high mortality of male zygotes or offspring) are much more likely to produce the desired population suppression. As described previously, late-stage juvenile mortality can also aid supression by maintaining strong density-dependent competition at juvenile stages.

The effect of gene drives on the reproductive interactions between individuals and its implications for gene drive spread are generally poorly understood. While viability of transgenic individuals under natural conditions has begun to receive some much-needed attention, more research will be required on the reproductive abilities of transgenics in natural populations.

**CONCLUSIONS:**

A large number of ecological factors will affect the fate of gene drives in natural populations. Our ability to accurately forecast the impact of gene drives on populations will depend upon how well we understand these ecological effects. The modeling work reviewed here shows that variation in ecological factors is likely to have an especially strong impact on the success of suppression drives compared to that of replacement drives.

As has been highlighted by other authors, self-sustaining, non-threshold gene drives should only be developed for applications where any consequences of unrestricted spread are viewed by society as acceptable because of the positive gains expected. In these cases, a better understanding of the pest population's ecology could assist in the development of gene drives most suited to a project's goals, and could guide efficient strategies for spatial and temporal field releases of the drives.

In cases where unrestricted spread of a gene drive is deemed societally unacceptable, gene drives with substantial release thresholds or other strong localizing mechanisms may provide a solution for a contained management of pest populations. However, the modeling work done to date indicates that it can sometimes be difficult to achieve *any* spatial spread for threshold drives, especially those with high release thresholds. Creating gene drives that can achieve the balance between ability to spread and to remain locally confined is a challenge that will require a better understanding of the match between the properties of the specific gene drives and the pest population's ecology in the targeted environment. Sensitivity analyses based on detailed and locally relevant models could be useful in guiding the direction and intensity of empirical research for measuring the relevant ecological parameters. This research could, in turn, help to build more robust models that are useful for regulatory decision making.

Breakthroughs in molecular biology are resulting in rapid progress toward genetically stable gene drives that could be released into natural populations. It is imperative that the ecological and population genetic research keep pace with the molecular work if we are to make informed decisions about release of gene drives to address the myriad conservation, agricultural and human health issues for which their use has been proposed.

**ACKNOWLEDGEMENTS:**

We are grateful to Austin Burt, Brandon Hollingsworth, Bruce Hay, Jackson Champer, Jennifer Baltzegar, Michael Vella, and Thomas Prowse for providing valuable feedback on a previous version of this manuscript. All authors were funded by NIH-NIAID 1R01AI139085-01.